\documentclass[a4paper,amssymb,amsmath,pra,twocolumn,floatfix,showpacs,nobalancelastpage,superscriptaddress]{revtex4-1}

\usepackage{enumerate,
fancyhdr,
amsmath,
amssymb,
graphicx,
color}
\newcommand{\ket}[1]{|#1 \rangle}
\newcommand{\bra}[1]{\langle #1|}

\DeclareMathOperator{\sinc}{sinc}

\newcommand{\Sigin}[0]{\Sigma^{\text{in}}}

\graphicspath{{./}}

\begin{document}

\title{The effect of magnetic impurity scattering on transport in topological insulators} %

\author{\firstname{Jesse} A. \surname{Vaitkus}}
\affiliation{Chemical and Quantum Physics, School of Science, RMIT University, Melbourne 3001, Australia}
\affiliation{ARC Centre of Excellence in Future Low-Energy Electronics Technologies}
\affiliation{HQS Quantum Simulations GmbH, 76187 Karlsruhe, Germany}

\author{\firstname{Cong Son} \surname{Ho}}
\affiliation{Chemical and Quantum Physics, School of Science, RMIT University, Melbourne 3001, Australia}
\affiliation{ARC Centre of Excellence in Future Low-Energy Electronics Technologies}

\author{\firstname{Jared} H. \surname{Cole}}
\email{jared.cole@rmit.edu.au}
\affiliation{Chemical and Quantum Physics, School of Science, RMIT University, Melbourne 3001, Australia}
\affiliation{ARC Centre of Excellence in Future Low-Energy Electronics Technologies}

\begin{abstract}\noindent 
Charge transport in topological insulators is primarily characterised by so-called topologically projected helical edge states, where charge carriers are correlated in spin and momentum. In principle, dissipation-less current can be carried by these edge states as backscattering from impurities and defects is suppressed as long as time-reversal symmetry is not broken. However, applied magnetic fields or underlying nuclear spin-defects in the substrate can break this time reversal symmetry. In particular, magnetic impurities lead to back-scattering by spin-flip processes. We have investigated the effects of point-wise magnetic impurities on the transport properties of helical edge states in the BHZ model using the Non-Equilibrium Green's Function formalism and compared the results to a semi-analytic approach. Using these techniques we study the influence of impurity strength and spin impurity polarization. We observe a secondary effect of defect-defect interaction that depends on the underlying material parameters which introduces a non-monotonic response of the conductance to defect density. This in turn suggests a qualitative difference in magneto-transport signatures in the dilute and high density spin impurity limits.
\end{abstract}

\maketitle

\section{Introduction}
Since their experimental observation in 2007 \cite{Konig2007}, topological insulators (TIs) have represented a paradigm shift in condensed matter physics. These materials exhibit novel surface state conduction that originates from the underlying topology character of the bulk material. These edge states have been shown to possess  spin-momentum locking, preventing back-scatter when time-reversal symmetry is preserved. These effects offer a new opportunity in device designs that utilize the underlying spin-texture of these edge states and present a new opportunity for spintronic technologies. In order for these so-called protected symmetries to exist, time-reversal symmetry must not be broken. Recent experiments have shown that nuclear spin states in the underlying substrate are found to affect transport by breaking this symmetry \cite{Tian2017,Liu2020}.

 
 
Theoretical work by several authors \cite{Bozkurt2018,Lunde2012,Buddhiraju2014} suggests that in the absence of strong nuclear dephasing, intrinsic nuclear spins may become polarized due to the flow of spin-polarised HESs in a TI. This results in little-to-no scattering once the nuclear spins are fully polarized. This polarization suggests that one could design a so-called spin-battery~\cite{Bozkurt2018} that stores a spin polarization that discharges when the edge states are ``turned off'', leading to questions of what its operational framework, efficacy, and efficiencies might look like. Recent work by Tian \textit{et al.}\cite{Tian2017} successfully demonstrated long-lived currents on the 40hr (4hr) time scale at 1.6K (45K) further stimulating these questions. More generally, the role of magnetic impurities, either from intrinsic spins or effective charge puddles is not yet well understood in TIs \cite{Liu2020,Zhao2019}.

In order to understand the interplay between underlying charge carriers, nuclear spins, and atom-impurity defects, it is useful to have spatially and energetically resolved spin--, charge--, and probability--densities. The Non-Equilibrium Green's Function (NEGF) formalism allows us to probe each of these quantities \cite{D2005}, allowing a fully spatially resolved simulation. In this paper we study the influence of point-like magnetic impurities on the conduction properties of the Bernevig, Hughes and Zhang (BHZ) model \cite{BHZ2006} using the NEGF formalism. Furthermore, we also support our numerical simulation with a semi-analytic analysis to describe the electron-defect interaction, where the back-scattering is allowed via spin-flip process. Our numerical and analytic methods show good qualitative agreement.

The paper is arranged as follows, in section \ref{sec:method} we introduce the simulation method including Hamiltonian and impurity models, then investigate the effects of impurity polarization and spatial position in section \ref{sec:imppol} and then the effects of impurity-impurity interaction in section \ref{sec:impimp}.

\section{Modelling magnetic impurity scattering in the BHZ model}\label{sec:method}
In this section we detail the models that we will be using for computing the transport response of the BHZ model in the presence of magnetic defects.

\subsection{BHZ Hamiltonian}
We use the quantum spin-hall (QSH) insulator model of Bernevig, Hughes and Zhang (BHZ)\cite{BHZ2006} derived originally for their work in HgTe/CdTe wells:
\begin{multline}
\mathcal{H}_\text{BHZ} = \epsilon_k \sigma_0 \tau_z + E_k \sigma_0 \tau_0 + \Delta_0 \sigma_y \tau_y \\ 
 + A_0 \left( k_x \sigma_z \tau_x  - k_y \sigma_0 \tau_y \right) 
\end{multline}
where $E_k = C_0 - D_0 (k_x^2 + k_y^2)$, $\epsilon_k = M_0 - B_0 (k_x^2 + k_y^2)$, $A_0$, $B_0$, $C_0$, $D_0$, $M_0$, $\Delta_0,$ are material parameters discussed below, and $\sigma$ and $\tau$ are Pauli matrices to denote spin and electron-hole degrees of freedom respectively. Making the canonical substitution of \mbox{$k_\alpha \rightarrow -i \partial/\partial \alpha$} and expanding we arrive at:
\begin{multline}\label{eq:BHZHam}
\mathcal{H}_\text{BHZ} = (M_0 + B_0 \nabla^2 ) \sigma_0 \tau_z + (C_0 + D_0 \nabla^2) \sigma_0 \tau_0 \\
 + \Delta_0 \sigma_y \tau_y - i A_0 \left( \frac{\partial}{\partial x} \sigma_z \tau_x  -  \frac{\partial}{\partial y} \sigma_0 \tau_y \right)
\end{multline}
where $\nabla^2$ is the Laplacian ($\frac{\partial^2}{\partial x^2} + \frac{\partial^2}{\partial y^2}$), $A_0$ in units of ($\mathrm{eV}$\AA) is the ``Dirac'' part of the equation, $B_0$ ($\mathrm{eV}$\AA$^2$) is the ``Schr\"odinger'' part of the equation, $C_0$ ($\mathrm{eV}$) is an energy offset, $D_0$ ($\mathrm{eV}$\AA$^2$) controls the asymmetry between valence and conduction bands. $M_0$ ($\mathrm{eV}$) controls the inversion region, (topologically interesting regions being $M_0 B_0 >0$), and also determines the character of the edge state at given $k_x$ \cite{Zhou2008,Shan2010}, whereas $\Delta_0$ controls the lowest-order bulk-inversion asymmetry (BIA). Given two well separated edge states, and $2 |\Delta_0| \ll A_0^2/|B_0|$, we obtain the effective edge Hamiltonian as
\begin{eqnarray}\label{eq:edgemode}
\mathcal{H}_\text{edge}=\epsilon_0+\hbar v_0 \sigma_z k,
\end{eqnarray}
which has a linear dispersion about the Dirac point $\epsilon_{k} = \epsilon_0 \pm \hbar v_0 k$, where $\epsilon_0=C_0 - M_0 D_0/B_0$, and $v_0 = \sqrt{B_0^2 - D_0^2} |A_0|/(\hbar B_0)$  \cite{Lunde2012} corresponding to the spin-polarized edge modes.

We set $C_0 = 0 \mathrm{~eV}$, \mbox{$D_0 = 0 \mathrm{~eV} $\AA$^2$}, such that we have a symmetric system centred at \mbox{$E = 0 \mathrm{~eV}$}. We set $\Delta_0 = 1~\mu\mathrm{eV}$, to serve as a method of lifting the degeneracy found between the two sub-blocks. Note that this term does not violate the $\mathbb{Z}_2$ symmetry class \cite{Konig2008}, but spin is no longer a conserved property \cite{Qi2011, Krueckl2011}. This judicious choice of $\Delta_0$ will make the magnitude of any new effects very small, but its purpose is solely to ease computational complexity found when dealing with degeneracies and does not reduce the generality of our results.

In HgTe/CdTe the experimentally observed $\Delta_0$ was approximately $1.6\mathrm{~meV}$, only one order of magnitude smaller than the band gap \cite{Konig2008}. However the effects we study here still apply qualitatively to such systems with much larger $\Delta_0$. In the absence of an asymmetry parameter $D_0$, the spatial and energetic scales of the BHZ model may be found by:
\begin{equation}
r_0 = \frac{2 |B_0|}{A_0}, \quad E_0 = \frac{A_0^2}{2 |B_0|}.
\end{equation}
One may then invert the length scale parameters to find the appropriate $A_0$ and $B_0$ parameters for a desired model:
\begin{equation}
A_0 = r_0 E_0, \quad  B_0= r_0^2 E_0/2.
\end{equation}
the inclusion of an asymmetry parameter, $D_0$, breaks the particle-hole symmetry and leads to competing energy and length scales which is not of interest to this work. 

\begin{figure*}\centering
	\includegraphics[]{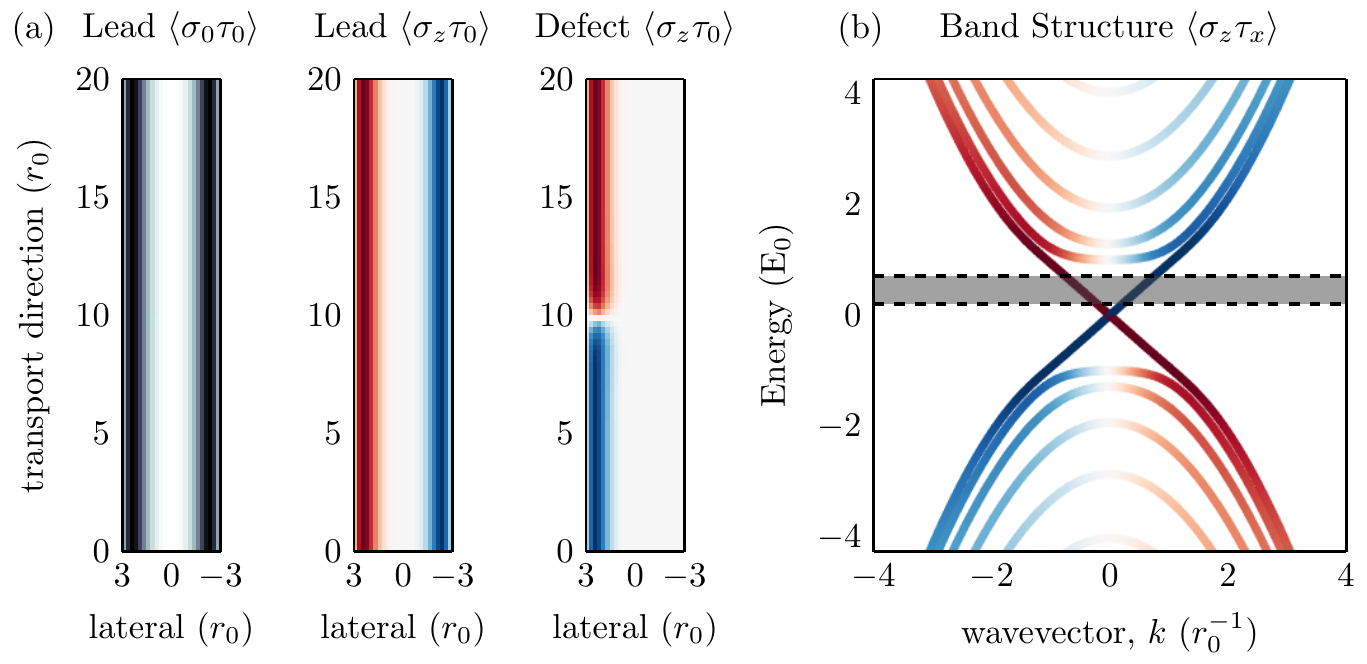}\vspace{0.5em} \\
	\includegraphics[]{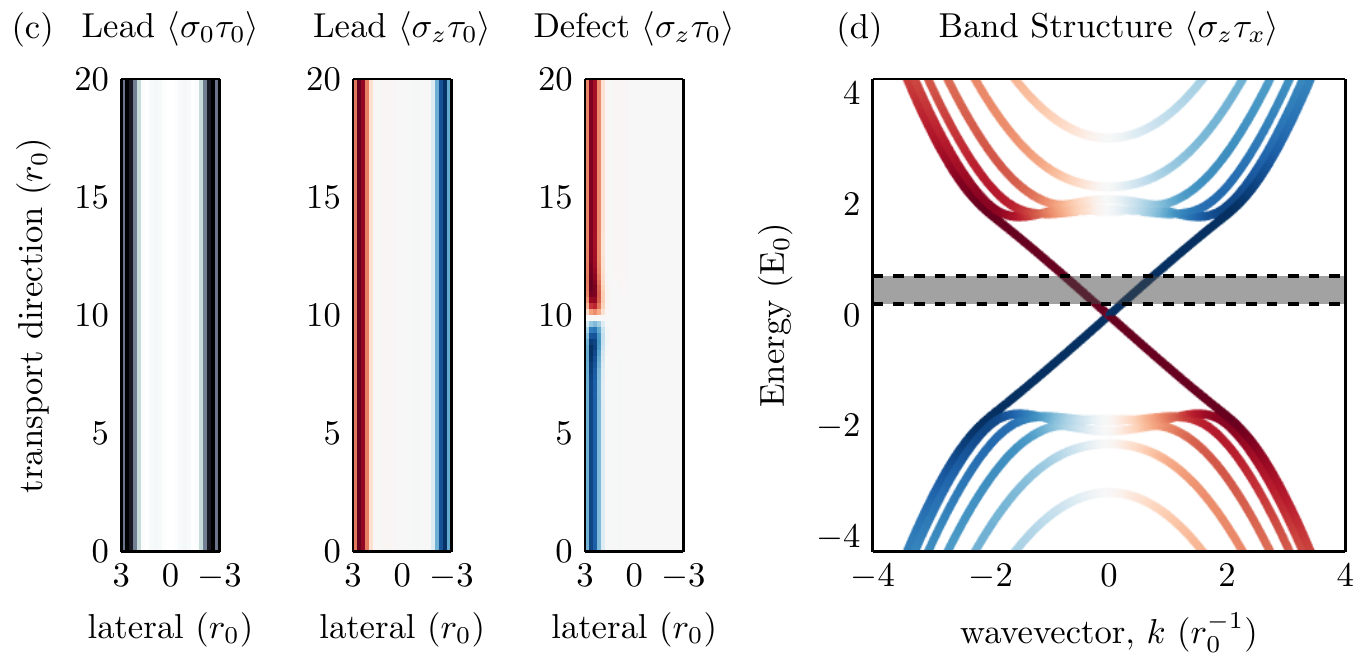}\vspace{1em}\\
	\hspace{3em} \includegraphics[]{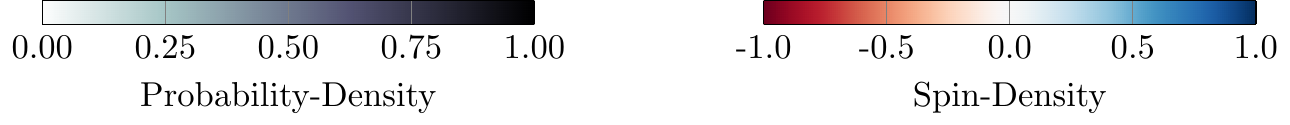}
	\caption{ (a,c) Local Density of States (LDOS; grayscale plot), Spin-projected charge density (SPCD; colour plot) for leads/defects, and (b,d) band structure for $M_0 = 1.0 E_0\text{ (a,b) and } 2.25 E_0$ (c,d) respectively. The bias window used in this work is shaded in the band structure. $M_0>A_0^2/(2B)$ causes an inversion of the bands, greater confinement towards the edges, and adds small extra nodes in the channel. For low $k$, these values of $M_0$ correspond to highly confined edge states \cite{Lu2012,Shan2010}. }
	\label{fig:varyM}
\end{figure*}

As we employ a real-space Greens function approach, we express the Hamiltonian on a grid using the finite difference method. This means that the results of our calculations can be expressed with respect to an arbitrary spatial scale, provided the relationship to the energy scale is defined. In everything that follows we choose $A_0 = 0.5\mathrm{~eV} r_0$, $B_0 = 0.25\mathrm{~eV} r_0^2$, giving a spatial scale $r_0$, corresponding to an energy scale of $E_0 = 500\mathrm{~meV}$. We vary $M_0$ from 1 to 2.25$E_0$, demonstrated in Fig.~\ref{fig:varyM}, noting that at $M_0 = A_0^2/(2|B_0|)=E_0$, the conventional band edge crosses over from direct to an indirect band gapped material.
In addition, at $M_0=E_0$, the bulk band gap is at its largest $\approx 2E_0$ comparable to state-of-the-art materials in this field \cite{Zhao2019}. 

In this work we leave $r_0$ as an arbitrary spatial scale, which may then be chosen for a given material. To give a sense of typical values for the length and energy scales, 
for a 7nm wide quantum well in HgTe, $r_0=3.76$nm and $E_0=96.8 \mathrm{meV}$\cite{Konig2008}. Another example is the quantum spin hall insulator WTe$_2$ which has a similar but not identical effective Hamiltonian\cite{Tang:2017,Qian2014}, which for our purposes we will approximate as isotropic to fit the form of Eq.~\ref{eq:BHZHam}. In this case, if we assume $A_0\approx2.5 \mathrm{eV\AA}$ and $|B_0|\approx34 \mathrm{eV\AA^2}$\cite{Qian2014}, we obtain $r_0\approx2.72$nm and $E_0\approx92 \mathrm{meV}$. For reference, in terms of Fermi velocity near the Dirac point HgTe/CdTe is approximately $3.7 \times 10^5 \mathrm{m/s}$~\cite{Lunde2012} and WTe$_2$ is approximately $3.9 \times 10^5 \mathrm{m/s}$~\cite{Qian2014}. 

Following the work of Lunde \textit{et al.} \cite{Lunde2012} it is suggested that if strong spin-polarization effects are to be observed, then the bias must be at least $V=5k_BT/e$, to this end we use biases ranging from $50~\mathrm{mV}$ to $250~\mathrm{mV}$, and therefore we model our devices to be at an operating temperature of $116$K, \textit{i.e.} $k_BT = 10~\mathrm{meV}$, approximately $1\%$ of the band gap. 

To express the BHZ Hamiltonian in real space, we utilize the second order 3--point finite difference method to discretize our derivative operators on a finite square lattice, where we set our discretisation constant (in both dimensions) to be 0.25~$r_0$. 



\subsection{NEGF transport model}

We use the Non-Equilibrium Green's function (NEGF) formalism \cite{D2005} to explicitly model transport in the presence of magnetic defects. The retarded Green's function matrix is given by:
\begin{equation}
G^r(E) = \left[ (E+i0^+) S - H - \Sigma^r \right]^{-1}  \label{eq:Greenfun}
\end{equation}
where $E$ is energy, $0^+$ is a small positive number to provide numerical stability, $S$ is the basis overlap matrix (equal to the identity in orthogonal bases and herein), $H$ is the Hamiltonian, $\Sigma^r$ is the sum of all retarded self-energy terms controlling the inflow and outflow of particles. In this work we will be using two sets of self-energy terms, one related to the leads as calculated using the iterative method of \cite{SSR1985,ONK+2010} and a second set related to spin-flip processes \cite{Yanik2007}, discussed in the following section. The distributions of input and output states are given by the in-- and out-- scattering functions $\Sigma^\text{in,out}$ respectively, in this work all \emph{leads} may be described using Fermi-Dirac distribution functions:
\begin{equation}
f_k(E) = \left[ \exp\left( \frac{E-\mu_k}{k_B T} \right) + 1 \right]^{-1}.
\end{equation}
The lead in-- and out-- scattering functions are given by weighted sum of the broadening matrices multiplied by their Fermi-Dirac distribution functions:
\begin{equation}
\Sigma^{\text{in}}_\text{leads} = \sum_{k} f_k \Gamma_k, \quad \Sigma^{\text{out}}_\text{leads} = \sum_{k} (1-f_k) \Gamma_k,
\end{equation}
where the $k^\text{th}$ broadening matrix $\Gamma_k$ is given by:
\begin{equation}
\Gamma_k = i( \Sigma_k - \Sigma_k^\dagger).
\end{equation}
One may then compute the electron and hole Green's functions \cite{D2005} by:
\begin{equation}
G^{n,p} = G \Sigma^\text{in,out} G^\dagger, \quad \Sigma^\text{in,out} = \Sigma^\text{in,out}_\text{leads} + \Sigma^\text{in,out}_\text{sf}
\end{equation}
where $\Sigma^\text{in,out}_\text{sf}$ is the in--(out--)scattering function due to spin-flip model discussed in the following section. It should be noted that electron and hole is a misnomer in this case, as these matrices represent the occupied and unoccupied states, sometimes referred to as the lesser and greater than functions ($-i G^{<} $ and $i G^{>}$ respectively) which are in fact super-positions of electron and hole states. Each of these Green's functions are then calculated by the recursive Green's function (RGF) approach~\cite{Anantram2008}.

To compute the expectation value of various operators with respect to the Green's functions, we first recall the periodic representation of the Hamiltonian:
\begin{equation}
H(k) = H_0 + H_1 e^{ika} + H_{-1} e^{-ika}  \label{eq:Hamk}
\end{equation}
where $H_0$ is the on-site matrix, $H_1=H_{-1}^\dagger$ is matrix connecting one unit cell to the next in the $k$--direction, and $a$ is the size of the unit cell. From (\ref{eq:Hamk}) we obtain the eigenproblem:
\begin{equation}
H(k) \ket{\Psi_j(k)} = E_j \ket{\Psi_j(k)}, 
\end{equation}
whose solutions form the bands in Fig.~\ref{fig:varyM}. Observables for the band structure are then obtained by computing the expectation value of the eigenfunction with respect to the operator:
\begin{equation}
\langle \sigma_a \tau_b  \rangle =  \bra{\Psi_j(k)} \sigma_a \tau_b \ket{\Psi_j(k)}.
\end{equation}
Similarly, the expectation value obtained from a Green's function is taken by computing the trace of an operator product with the Green's function:
\begin{equation}
\langle \sigma_a \tau_b \rangle = \frac{1}{2 \pi} \text{Tr}[ X (\sigma_a \tau_b) ], \quad X = A, G^n, G^p. \label{eq:operator}
\end{equation}
With common examples being $\text{Tr}[A (\sigma_0 \tau_0)]$ for local density of states (LDOS), and $\text{Tr}[G^n (\sigma_z \tau_0)]$ for spin-projected charge density (SPCD). The band structure, local density of states and the spin-projected charge density for the lead and defect contributions are presented in Fig.~\ref{fig:varyM}.

Additionally, the standard current operator \cite{D2005} through the $i^\text{th}$ lead:
\begin{equation}
I_i = -\frac{e}{h} \int_{-\infty}^{\infty} \text{Tr}[ (\Sigma_i^\text{in} A - \Gamma_i G^n ) ] dE, \label{eq:oldcurrent}
\end{equation}
must be modified to account that the electrons and holes travelling in the same direction each carry opposite currents, and therefore we modify (\ref{eq:oldcurrent}) to handle particle--hole symmetries\cite{Sriram2019}:
\begin{equation}
I_i = -\frac{e}{h} \int_{-\infty}^{\infty} \text{Tr}[ (\Sigma_i^\text{in} A - \Gamma_i G^n ) \sigma_0 \tau_z  ] dE
\end{equation}
alternatively this may be written as a difference between the partial traces across the electron and hole subspaces.

\subsection{Effective scattering model}\label{spinflip}
Before we compute numerical results using NEGF, it is illustrative to first consider an effective scattering model in which we consider a reduced system containing solely an electron and spin-defect described by the following Hamiltonian:
\begin{eqnarray}
H_{\rm{eff}}=H_\text{edge}+H_i+H_\text{int}.
\end{eqnarray}
In the above $H_\text{edge}= \hbar v_0k_x\sigma_z$ describes electrons in the edge channel along the $x$--direction given by Eq. (\eqref{eq:edgemode}). $H_i$ is the unperturbed Hamiltonian of the defect spin. In the absence of external magnetic field, $H_i$ describes degenerate states of the defect spin and we set it as a constant $H_i=0$ without loss of the generality. The last term is the interaction between electron-spin and defect-spin given as \cite{Gao2009,Culcer2011,Zhu2011,Jiang2011a}
\begin{eqnarray}
H_\text{int}=F(x)[J_z\sigma_z S_z+J_x(\sigma_-S_+ + \sigma_+ S_-)],
\end{eqnarray}
where $F(x)=\Theta(w_x/2-|x|)$ is a range function describing the effective size of the scattering region, $S$ is the defect spin operator, and $\sigma_{\pm}=(\sigma_x\pm i\sigma_y)/2$. We note here that the dipolar coupling between the defect spins is much weaker than the hyperfine interaction \cite{Paget1977,Hsu2018}, and thus can be ignored. 

In the presence of electron-defect interaction, the time-reversal symmetry is broken and allows back-scattering via the spin-flip process. For the simplicity, we consider the case of spin-1/2 scatterer, which is realistic in a HgTe quantum well \cite{Hsu2018,Lunde2013}. We can construct four basis vectors as $\ket{\sigma_z,S_z} = \ket{\! \! \uparrow \uparrow}, \ket{\! \! \uparrow \downarrow}, \ket{\! \! \downarrow\uparrow}, \ket{\! \! \downarrow \downarrow}$, and the total Hamiltonian is cast into:
\begin{equation}\hspace{-1.0em}
H_{\rm{eff}}= \resizebox{.4\textwidth}{!}{$ \left(\begin{array}{cccc}
	\hbar v_0 k+J_z&0&0&0\\
	0&\hbar v_0 k - J_z&J_x&0\\
	0&J_x&-\hbar v_0 k-J_z&0\\
	0&0&0&-\hbar v_0 k+J_z\end{array}\right).$}
\end{equation}
The middle block corresponds to the spin-flip scattering (\textit{i.e.} the total spin of electron and impurity is conserved ${J=\langle\sigma_s+S_z\rangle=0}$), is the subdomain of interest. This reduced system may then be explicitly written as \cite{Russo2018}:
\begin{eqnarray}
H_\text{sf}=\left(\begin{array}{cc}
k\hbar v_0-J_z&J_x\\
J_x&-\hbar v_0 k-J_z\end{array}\right).
\end{eqnarray}
 The eigenstates are given by:
\begin{align}
\psi_+(q)=&\left(\begin{array}{c}
\ \cos \theta/2\ \\
\ \sin \theta/2 \ \end{array}\right)e^{iqx}, \\ \psi_-(q)=&\left(\begin{array}{c}
-\sin \theta/2\\
\cos \theta/2\end{array}\right)e^{iqx},
\end{align}
where we have defined angles ${\sin\theta\! =\! J_x/\sqrt{J_x^2+(\hbar v_0 q)^2}}$, ${\cos\theta \! = \! \hbar v_0 q/\sqrt{J_x^2+(\hbar v_0 q)^2}}$, and the corresponding energies are ${E_\pm=-J_z\pm\sqrt{J_x^2+(\hbar v_0 q)^2}}$. We now investigate the spin-flip properties in the following subsection.

\subsubsection{Single defect}
Using the wavefunctions of the underlying edge states we consider the elastic scattering of an electron with energy $E>0$ incident from the left of the defect. Outside the scattering region, the electron wavefunctions on the left and right sides of the defect are 
\begin{eqnarray}
\psi_L= \left(\begin{array}{c}
\alpha e^{ikx}\\
r\alpha e^{-ikx}\end{array}\right), \ \ \psi_R= \left(\begin{array}{c}
t\alpha e^{ikx}\\
0\end{array}\right)
\end{eqnarray}
with $r,t$ being reflection and transmission coefficients respectively, and the momentum is given by $k=E/\hbar v_0$.

At the same time, elastic scattering requires that the energy of electron in the scattering region is also positive, which is given by $E_+>0$.  In this case, the wavefunction of electron in the scattering region is given by
\begin{align}\label{phic}
\psi_c =& c_1\psi_+(q)e^{iqx}+c_2\psi_+(-q)e^{-iqx}.
\end{align}
 with $q=\sqrt{(E+J_z)^2-J_x^2}/(\hbar v_0)$.

Imposing the continuity relation at the left and right sides of the scattering region, ${\psi_L(-w_x/2)=\psi_c(-w_x/2)}$, ${\psi_R(w_x/2)=\psi_c(w_x/2)}$, we obtain the transmission and reflection coefficients as
\begin{gather}
T=|t|^2=\frac{\cos^2\theta}{1- \sin^2\theta \cos^2{qw_x}},\\
R=|r|^2=\frac{\sin^2qw_x}{\sin^2qw_x+\cot^2\theta},
\end{gather}
which satisfy the relation $R+T=1$. 

Having obtained the transmission probability, the conductance is simply derived as $\sigma=G_0\langle T\rangle$, where $\langle ...\rangle$ represents taking average value over all occupied states. At low temperature, the conductance decrease in the 1D edge channel is:
\begin{eqnarray}\label{sfG}
\delta \sigma=G_0\langle \delta T\rangle=G_0\int_0^{E_\text{f}}dE \delta T(E).
\end{eqnarray}

For point-like defect, we can assume that the scattering region is small (weak interaction) \textit{i.e.} $kw_x \ll 1$, then to the leading order in $w$, the transmission simplifies to
\begin{eqnarray}\label{T1}
T&&\approx 1-q^2w_x^2\tan^2\theta.
\end{eqnarray}
Noting that $\tan\theta=J_x/(\hbar v_0 q)$, we obtain the change in the transmission coefficient due to the scattering as
\begin{eqnarray}
\delta T=R=\left(\frac{w_x J_x}{\hbar v_0}\right)^2. \label{eq:deltaT}
\end{eqnarray}
From this expression we see that to lowest order, the reflection should scale with the square of the effective defect length, $J_x^2$, and inversely with the Fermi velocity squared. Surprisingly, the reflection coefficient is independent of the energy and momentum of the incident electron, and it is entirely expressed as a function of the defect parameters (defect size, hyperfine interaction) and the edge state Fermi velocity. 

\subsubsection{Multiple defects}
Similarly, we can extend the above arguments to the scattering problem with $N_I$ defects. For simplicity, we assume that the defects are linearly distributed along the edge channel with equal separation $l$. The transmission probability for $N_I$ defects can be obtained by
\begin{eqnarray}\label{multiT}
T_{N_I}\approx 1- \delta T\left(\frac{\sin N_I kl}{\sin kl}\right)^2,
\end{eqnarray}
with $\delta T$ given in (\ref{eq:deltaT}). In the presence of multiple defects, the net change in transmission coefficient is not simply additive. Consequently the effect of multiple defects in spin-flip scattering is non-trivial, as we will also see in the NEGF analysis.

Nevertheless, when the separation between defects is large \textit{i.e.} $l\gg 1/k_\text{f}$, as is the case for low defect density, Eq. \ref{multiT} is simplified to
\begin{eqnarray}
T_{N_I}=1-N_I\delta T,
\end{eqnarray}
which gives an approximate scaling of the multiple defects. This scaling indicates a trivial additive effect and can be understood as a consequence of the decoupling between well-separated defects, where the net transmission coefficient is just the product of individual coefficients at each site \textit{i.e.} $T_N = T_1^N = (1-\delta T)^N \approx 1-N\delta T$. It is worth noting here that this additive limit is what was assumed (for example) in Ref.~\cite{Bozkurt2018} resulting in a trivial dependence on the density of spin impurities.

\subsubsection{Dependence on lateral position of defects}\label{latfun}
For the previous derivation we assumed that defects act as one-dimensional scatterers, and computed the reflection coefficient assuming such. In fact the coupling of spin defects to the edge state transport will depend on their spatial positions due to the spatial variation of the edge modes.

Assuming well-separated edges and vanishingly small $\Delta_0$, the wavefunction is given by a lateral localized function $f(y)$ of the form \cite{Zhou2008}:
\begin{eqnarray}\label{transfun}
f(y)= \sqrt{\frac{2\lambda_1\lambda_2(\lambda_1+\lambda_2)}{(\lambda_1-\lambda_2)^2}} \left(e^{-\lambda_1 y}-e^{-\lambda_2 y}\right),
\end{eqnarray}
where parameters $\lambda_{1,2}$ are determined near the Dirac point by:
\begin{eqnarray}
\lambda_{1,2}=\frac{1}{\sqrt{B_0^2-D_0^2}}\left(\frac{|A_0|}{2}\pm\sqrt{W_0}\right),
\end{eqnarray}
with $W_0=A_0^2/4-M_0(B_0^2-D_0^2)/B_0$.

From the wavefunction in (\ref{transfun}), we can determine the effective width of the helical edge states as well as the lateral position for maximum electron density. Explicitly, solving $f'(y)=0$ we obtain:
\begin{eqnarray}
y_\text{peak}=\frac{\log{\lambda_1}-\log{\lambda_2}}{\lambda_1-\lambda_2}.
\end{eqnarray}

For a defect placed at $y=y_i$, the average value of the conductance decrease across the lateral direction is:
\begin{align}
\langle \delta G \rangle =& \frac{1}{L_y}\int_{L_y} \ \delta G |f(y)|^2 g(y) dy   \nonumber \\
=& \frac{1}{L_y}\int_{L_y} \ \delta G |f(y)|^2 \Theta(w_y/2-|y-y_i|) dy  \nonumber \\
\approx& \delta G \frac{w_y}{L_y} |f(y_i)|^2.
\end{align}
where $\delta G$ is the proportional conductance decrease for a given event (assumed independent) and $g(y)$ is the spatial distribution of defects. Here for simplicity we assume the spatial distribution of the defect is given by a $\Theta$-function which defines a constant region of width $w_y$ where the defect is ``turned on''. From this, we can see that the true defect reduction is:
\begin{eqnarray}
\delta T=R= \frac{w_y}{L_y} \left(\frac{w_x  J_x |f(y_i)|}{\hbar v_0}\right)^2. \label{eq:deltaT2}
\end{eqnarray}

From (\ref{eq:deltaT2}) we see that the conductance decrease is directly proportional to the underlying probability density of the carrier wavefunctions. As a consequence, defects in the bulk will have only small contributions to the scattering process and transport in 2D TI films.


\subsection{Spin-flip model NEGF}
Now we introduce the mechanism of spin-flipping in our NEGF approach by utilizing a spin-flip self-energy based on the derivation by Yanik \textit{et al.}\cite{Yanik2007} extended to support polarized nuclear spin impurities. Firstly, we consider the nuclear spins as a bath with density matrix:
\begin{equation}
\rho = \left( \begin{array}{cc}
F_u & \Delta \\
\Delta^* & F_d 
\end{array} \right)
\end{equation}
where $F_{u,d}$ are the fraction of spins that are up (down), \mbox{$F_u + F_d = 1$}, and $\Delta$ is the coherence between these populations. For consistency with the derivation of Yanik we keep the $F_u, F_d$ nomenclature, however, one may transform this expression into magnetization $\langle m \rangle$ by:
\begin{equation}
F_{u,d} = \frac{1}{2}\left( 1 \pm \langle m \rangle  \right),
\end{equation}
by noting that $\langle m \rangle = F_u-F_d$. By explicitly assuming that the energy required for a spin-flip is elastic, one arrives at the following fourth order tensor relations for the in-- and out--scattering matrices, $\Sigma^\text{in,out}$:
\begin{multline}
\left[ \begin{array}{c} (\Sigma^{\text{in,out}}_{\uparrow \uparrow})_{jj} \\
(\Sigma^{\text{in,out}}_{\downarrow \downarrow})_{jj} \\ 
(\Sigma^{\text{in,out}}_{\uparrow \downarrow})_{jj} \\ 
(\Sigma^{\text{in,out}}_{\downarrow \uparrow})_{jj}  \end{array} \right ]
= K \left[ \begin{array}{c} (G^{n,p}_{\uparrow \uparrow})_{jj} \\
(G^{n,p}_{\downarrow \downarrow})_{jj} \\ 
(G^{n,p}_{\uparrow \downarrow})_{jj} \\ 
(G^{n,p}_{\downarrow \uparrow})_{jj}  \end{array} \right ], \\
K = \langle J^2 \rangle N_I \left[
\begin{array}{cccc}
1/4 & F_{u,d} & -\frac{\Delta ^*}{2} & -\frac{\Delta}{2} \\
F_{d,u} & 1/4 & -\frac{\Delta ^*}{2} & -\frac{\Delta}{2} \\
\frac{\Delta}{2} & \frac{\Delta}{2} & -1/4 & 0 \\
\frac{\Delta ^*}{2} & \frac{\Delta ^*}{2} & 0 & -1/4
\end{array}
\right] \label{eq:tensor}
\end{multline}
where $\langle J^2 \rangle$ is the effective squared coupling rate after averaging over $N_I$ impurities and energy, suppressing the $x$ subscript for brevity and noting that $j$ is the lattice site. Throughout we use $\langle J^2 \rangle N_I = 30\mathrm{meV}^2$. In general, each parameter in the model can be considered to be spatially varying however we suppress spatial variation in what follows without loss of generality. Explicitly evaluating (\ref{eq:tensor}) gives:
\begin{multline}
\Sigma^\text{in} / J^{2} = \left(
\begin{array}{cc}
F_u G^n_{\downarrow \downarrow} & 0 \\
0 &  F_d G^n_{\uparrow \uparrow}
\end{array}
\right) + 
\frac{1}{4} \left(
\begin{array}{cc}
G^n_{\uparrow \uparrow}  & - G^n_{\uparrow \downarrow} \\
- G^n_{\downarrow \uparrow} & G^n_{\downarrow \downarrow}
\end{array}
\right) \\ + 
\frac{1}{2} \left(
\begin{array}{cc}
-G^n_{\uparrow \downarrow} \Delta ^* - G^n_{\downarrow \uparrow} \Delta & (G^n_{\downarrow \downarrow} + G^n_{\uparrow \uparrow}) \Delta \\
(G^n_{\downarrow \downarrow} + G^n_{\uparrow \uparrow}) \Delta ^*  & - G^n_{\uparrow \downarrow} \Delta ^* - G^n_{\downarrow \uparrow} \Delta \end{array}
\right)
\end{multline}
where we have absorbed $N_I$ into $\langle J^2 \rangle$ and dropped the angled brackets for brevity, $J^2 = N_I \langle J^2 \rangle$. We may repeat this similarly for $\Sigma^\text{out}$, and noting that \mbox{$\Gamma = \Sigma^\text{in} + \Sigma^\text{out}$}, and \mbox{$G^p= A - G^n$}, we arrive at the broadening matrix due to spin flips:
\begin{multline}
\Gamma_\text{sf} / J^{2} = 
(F_u-F_d) \left(
\begin{array}{cc}
G^n_{\downarrow \downarrow} & 0 \\
0 &  -G^n_{\uparrow \uparrow}
\end{array}
\right) \\ + \left(
\begin{array}{cc}
F_d A_{\downarrow \downarrow} & 0 \\
0 &  F_u A_{\uparrow \uparrow}
\end{array}
\right) 
+ \frac{1}{4} \left(
\begin{array}{cc}
A_{\uparrow \uparrow}  & - A_{\uparrow \downarrow} \\
- A_{\downarrow \uparrow} & A_{\downarrow \downarrow}
\end{array}
\right)  \\ - 
\frac{1}{2} \left(
\begin{array}{cc}
A_{\uparrow \downarrow} \Delta ^* + A_{\downarrow \uparrow} \Delta & -(A_{\downarrow \downarrow} + A_{\uparrow \uparrow}) \Delta \\
-(A_{\downarrow \downarrow} + A_{\uparrow \uparrow}) \Delta ^*  & A_{\uparrow \downarrow} \Delta ^* + A_{\downarrow \uparrow} \Delta \end{array}
\right),
\end{multline}
where the first two terms correspond to spin-flip processes and the remaining to spin-dephasing and bath coherence processes respectively, therefore this model involves both spin-conserving and non-spin-conserving processes. In \mbox{Yanik \textit{et al.}} spin-dephasing processes were omitted, as it was assumed they did not contribute significantly to the observables being studied \cite{Yanik2007}. In this work we found their numerical inclusion to be sufficiently inexpensive, and therefore we can determine if these effects were in fact insignificant, which will be discussed in section \ref{sec:imppol}. We assume that the bath of nuclear spins is in an incoherent mixture at all times and therefore set $\Delta=0$. 

To obtain the self-energy of a given broadening matrix, one computes the Hilbert transform:
\begin{equation}
\Sigma^r(E) = - \frac{i \Gamma(E)}{2}  +  \frac{1}{\pi} \int \frac{\Gamma(E^\prime)/2}{E-E^\prime + i0^+} dE^\prime
\end{equation}
often written as the principle part (about the divergence), we add a finite imaginary term to avoid this divergence. Note that for pure dephasing, one may set $F_u=F_d$, and compute the spin-flip self-energies without involving the Hilbert transform as the spectral function is unambiguously related to the retarded Green's function; $\Gamma \rightarrow  \Sigma^r, A \rightarrow G^r$, and is given by
\begin{equation}
\Sigma^r_\text{sf}/J^2 = \frac{1}{2} \left(
\begin{array}{cc}
G^{r}_{\downarrow \downarrow} & 0 \\
0 &  G^{r}_{\uparrow \uparrow}
\end{array}
\right) + \\
\frac{1}{4} \left(
\begin{array}{cc}
G^{r}_{\uparrow \uparrow}  & - G^{r}_{\uparrow \downarrow} \\
- G^{r}_{\downarrow \uparrow} & G^{r}_{\downarrow \downarrow}
\end{array}
\right).
\end{equation}

\begin{figure*}[th!]\centering
	\includegraphics[]{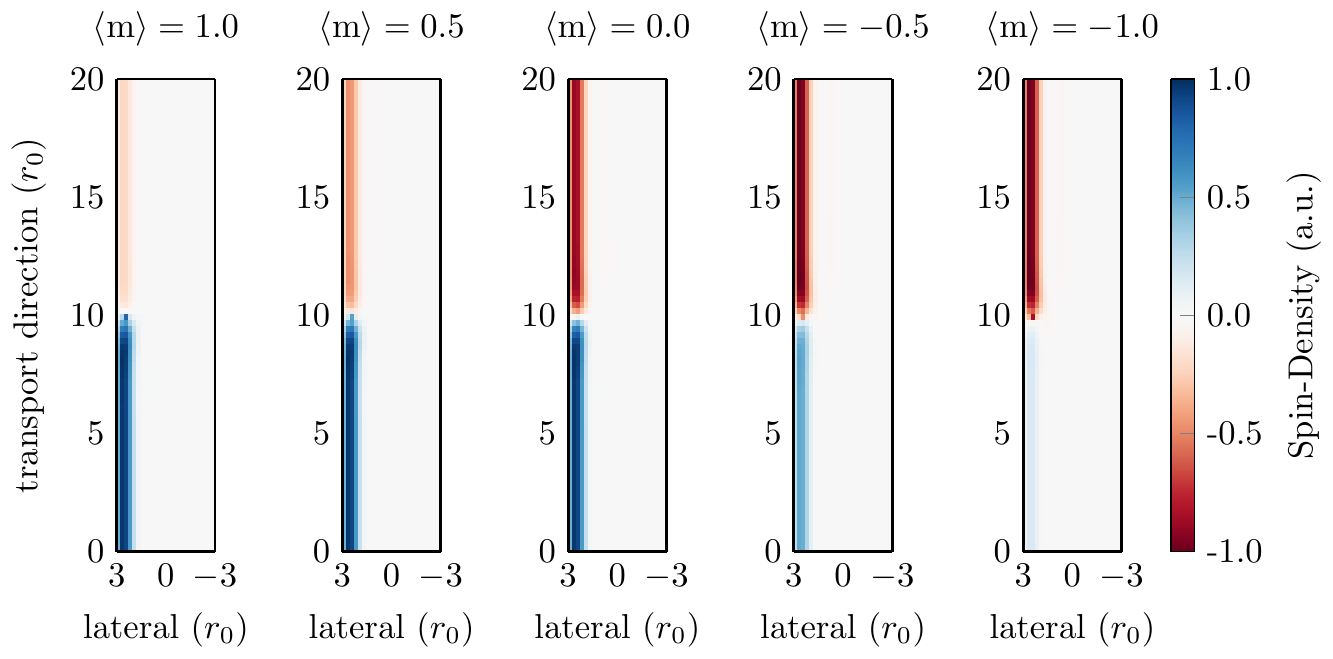}
	\caption{Current contribution due to impurity scattering as a function of defect magnetization. As the magnetization is shifted from up to down, the scatter current contribution also smoothly shifts from one direction to the other. For the scattered carriers, negative spin flow upwards and positive spin flow downwards due to spin-momentum locking.  Note that at total polarization $\langle m \rangle = \pm 1$, some charge is re-injected with the same spin, corresponding to the spin-dephasing processes. Data have been rescaled by their extrema such that they fit on the same colour-scale. } 
	\label{fig:reinjected}
\end{figure*}

However, for any other polarization, such a statement cannot be made about the electron/hole Green's functions and we must evaluate the Hilbert transform explicitly. For more discussion about how we compute the Hilbert transform efficiently, see Appendix~\ref{App:Hilbert}. Explicitly defining $H[f(x)]$ to be the Hilbert Transform of $f(x)$, we arrive at the self-energy:
\begin{multline}
\Sigma^{r}_\text{sf}/J^2 = \left(
\begin{array}{cc}
F_d G^{r}_{\downarrow \downarrow} & 0 \\
0 &  F_u G^{r}_{\uparrow \uparrow}
\end{array}
\right) +
\frac{1}{4} \left(
\begin{array}{cc}
G^{r}_{\uparrow \uparrow}  & - G^{r}_{\uparrow \downarrow} \\
- G^{r}_{\downarrow \uparrow} & G^{r}_{\downarrow \downarrow}
\end{array}
\right) \\ 
-\frac{1}{2}(F_u-F_d)\left(
\begin{array}{cc}
+ i G^n_{\downarrow \downarrow} -  H[G^n_{\downarrow \downarrow}]  & 0 \\
0 & - i G^n_{\uparrow \uparrow} +  H[G^n_{\uparrow \uparrow}]
\end{array}
\right) \label{eq:sfself}
\end{multline}

In previous works \cite{VC2018,Greck2015} it was shown that for certain models the in-scattering function may be replaced with some simpler form corresponding to B\"uttiker probes. We have found in this work that simple replacement with a Fermi-Dirac distribution (or sum thereof) --- though able to reproduce effects with the correct order of magnitude --- was insufficient to fully describe the physics related to magnetization, and led to an additional resistance term for collinear spins, in contradiction to the full in-scattering function. Therefore we use the analytic in-scattering function throughout.


\section{Dependence on impurity polarization}\label{sec:imppol}

To understand how a magnetic impurity disrupts a HES, we isolate the contribution of electron Green's function corresponding to the impurity in-scattering function:
\begin{equation}
	G^n_\text{sf} = G \Sigin_\text{sf} G^\dagger
\end{equation}
and compute its observables using (\ref{eq:operator}). The partial density obtained can be thought of as a fictitious lead that absorbs a charge and re-injects it with or without a spin-flip. This re-injected current mimics the expected behaviour of spin-dependent scattering on the edge-state conduction with spins of opposite sign propagating in the opposite direction. 

For all the data presented we discuss the \textit{decrease} in conductance as $d\sigma$, which we define as
\begin{equation}
\sigma_\text{pristine} - \sigma_\text{actual} = d\sigma_\text{defect},
\end{equation}
and present in units of the quantum conductance \mbox{$G_0=e^2/h$}. Note that for a pristine QSH TI, the conductance is exactly $2 G_0$ originating from the two helical edge states, each carrying one conductance. In this work, we place defects only in one of the two edge states at a given time.

Similarly for the change in current, we define $dI$:
\begin{equation}
I_\text{pristine} - I_\text{actual} = dI_\text{defect}
\end{equation}
which we present in units of $I_0$, the current of a single channel at 1 volt ($I_0 = e^2 \mathrm{V}/h$). For example at a bias of $V=250\mathrm{mV}$ the pristine current would be \mbox{$I_\text{pristine}=0.50 I_0 $}.

\begin{figure}\centering
	\includegraphics[]{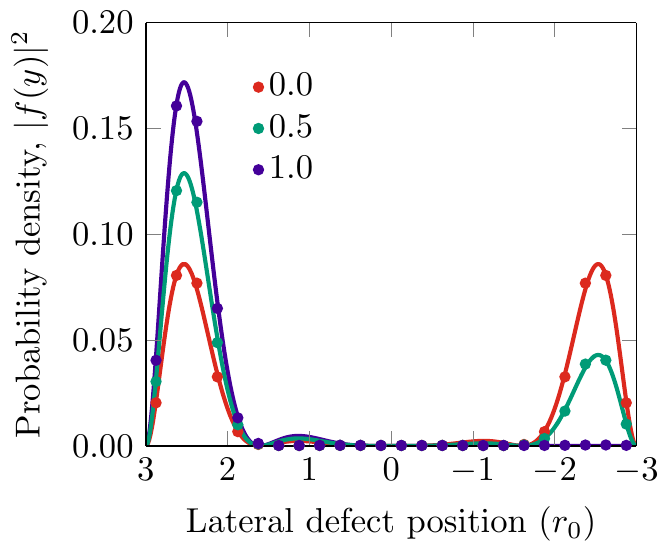}
	\caption{The defect scattering contribution to conduction as a function of position depends directly on the spatial distribution of the edge state. Dots are obtained by scaling the conductance decrease obtained by NEGF result by the factors presented in (\ref{eq:deltaT2}), showing the spatial dependence, for $M_0=2E_0$. This can be compared directly to the probability density of carriers (solid lines), with a trivial scale factor to match the conductance change. The ratios of each magnetization ($\langle m \rangle=0.0,0.5,1.0$) are $\mbox{2:2}$, $\mbox{3:1}$, and $\mbox{4:0}$ respectively, demonstrating the linear relationship between polarization and strength; collinear spins do not negatively affect transport. Note that negative magnetizations would be the same figure, but with the lateral defect position axis reversed.}
	\label{fig:conductancechange}
\end{figure}

In Fig.~\ref{fig:reinjected} we demonstrate the scattered spin-current as a function of the defect magnetization. As the defect polarization is smoothly transitioned from one orientation to the other, so too is the scattered current contribution. Note that even at maximal polarization in one direction ($\langle m \rangle = \pm 1$) some charge is scattered with spin of the opposite polarization, originating from the spin-dephasing term. This current does not negatively affect transport, the conductance change due to a collinear magnetic impurity is nil (see Fig.~\ref{fig:conductancechange}). As such, previous comments about its ability to be removed are seemingly well-founded \cite{Yanik2007}.

In Fig.~\ref{fig:conductancechange} we demonstrate the conductance change as a function of magnetic impurity position in the channel. Motivated by the derivation in (\ref{eq:deltaT2}) we normalize our result by the dimensionless factor $(w_y/L_y)(w_x J/\hbar v_0)^2$, where $w_x (w_y)$ is the width of the defect in the transport (lateral) direction, $J$ is the coupling strength and $\hbar v_0$ is the Fermi velocity. The conductance change is therefore directly proportional to the underlying probability density of the current carrying states.

As can be seen in Fig.~\ref{fig:conductancechange}, the change in conductance is highly dependent on the underlying carrier spin-density. When a defect is placed in higher density regions, the decrease in conductance is stronger. The region of high electron density can be determined by considering the transverse wavefunction $f(y)$, which has a localization peak at $y_\text{peak}$, as discussed in section~\ref{latfun}. Using the corresponding Hamiltonian parameters, we obtain $y_\text{peak}\approx 2.34~r_0$, which is consistent with the spatially-resolved NEGF method depicted in Fig.~\ref{fig:conductancechange} whose maximum density is $y = 2.375r_0 \pm 0.125 r_0$.

For a fixed density, the effect is linearly dependent on the underlying polarization between the defect and HES, being maximized when they are anti-parallel and minimized when they are parallel.  This is best observed by noting that the total average conductance decrease across the left and right edge states is \textit{constant} with polarization, being given by the ratios $\mbox{2:2}$, $\mbox{3:1}$, and $\mbox{4:0}$ for $\langle m \rangle = 0.0, 0.5, \text{ and } 1.0$ respectively. This is consistent with the spin-flip probability being proportional to both the charge-density at the impurity site and the impurity's polarization. Lastly, we note that the negative magnetizations ($\langle m \rangle=-0.5,-1.0$) would be the same as Fig.~\ref{fig:conductancechange}, but with the lateral defect position axis reversed. 

\begin{figure}\centering
	\includegraphics[]{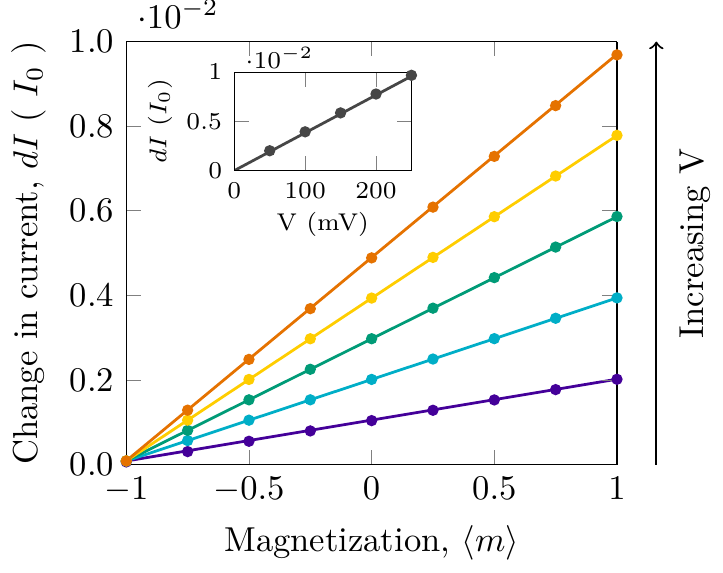} 
	\caption{The decrease in current as a function of magnetization at given biases. $I_0$ is the current of a single channel at 1 volt ($I_0 = e^2 \mathrm{V}/h$). Lines are given by a single bilinear fit in $\langle m\rangle$ and $V$ to the entire data set, demonstrating the linear sensitivity to both parameters. Bias increases upwards with the minimum and maximum biases being 50mV and 250mV respectively. Shown inset is the bias dependence when the defect is maximally polarized. To generalise our result we have normalized by the dimensionless factor $(w_x J /\hbar v_0)^2$ discussed in the text.}
	\label{fig:dImagnetization}  
\end{figure}

To further examine the functional dependence we place our defect at $y=2.375~r_0$,  corresponding to the maximum conductance change observed and plot it as a function of magnetization and conductance in Fig.~\ref{fig:dImagnetization}. Once again we see that when the defect magnetization axis is aligned with the dominant carrier, no response is observed, and when they are opposite, the effect is maximized. Furthermore the effect is linear in applied current, where the total response was found to be very well described by a bilinear fit over magnetization and bias suggesting that again, the conductance decrease linearly depends on the polarization of the defect, as is expected given the linear dependence of (\ref{eq:sfself}) on $F_u$ and $F_d$.

\subsection{Scattering-induced defect magnetization} 
The correlation between conductance and defect magnetization is consistent with spin-charge conversion induced by the helical edge state \cite{Tian2017}. The underlying physics can be understood in an intuitive way using the spin-flip scattering as discussed in the previous section. In this process, both electron and defect acquire non-equilibrium polarization pertaining to the conservation of the total spin. In the case of a single defect, the polarization of the defect is determined by $m=\mathrm{Tr}[S_z \rho_\text{tot}]$, where $\rho_\text{tot}=|\psi_c\rangle\langle\psi_c|$ is the density matrix, with $\psi_c$ given in Eq. (\ref{phic}), see Appendix~\ref{sfM} for full details. Explicitly,
\begin{eqnarray}
m=\frac{\cos^2\theta}{ \sin^2\theta \cos^2{qw_x} - 1},
\end{eqnarray}
which is exactly the expression of the transmission coefficient in Eq. (\ref{T1}) with an opposite sign, i.e., $m=-T$. Notice the relation between the conductance and the transmission coefficient, we can derive the linear relation between conductance change to the defect magnetization as $\delta \sigma\propto G_0(1+\langle m\rangle )$.

\subsection{Steady state magnetization}
So far, we consider the polarization induced by an electron. If there are $n_e$ non-interacting electrons in the spin-flip scattering, the total magnetization of the defect will be $ \mathcal{M}=\sum\delta m=n_e m$, from which we can then determine the magnetization rate as:
\begin{eqnarray}
\frac{d\mathcal{M}}{dt}=j_e m - \frac{\mathcal{M}}{\tau_{r}},
\end{eqnarray}
where $j_e=dn_e/dt$ is the charge current, and  $\tau_r$ is a phenomenological relaxation term \cite{Lunde2012}. In steady state, the magnetization is therefore obtained as
\begin{eqnarray}
\mathcal{M}_{st}=j_e\tau_r m,
\end{eqnarray}
which is consistent with previous work \cite{Lunde2012}. The above equation represents the current-induced polarization in the topological edge states. Conversely, a spin polarized defect can generate a charge current $I\propto d\mathcal{M}/dt$ via the  magnetization relaxation. In general a fully self-consistent solution of both electron and impurity spins is required~\cite{Buddhiraju2014} however here we focus on the effect of steady state spin impurities on the edge state conduction.





In the following section we discuss the material parameter $M_0$ and the signatures of impurity-impurity interactions. 

\section{Impurity proximity effects}\label{sec:impimp}

\begin{figure}\centering
	\includegraphics{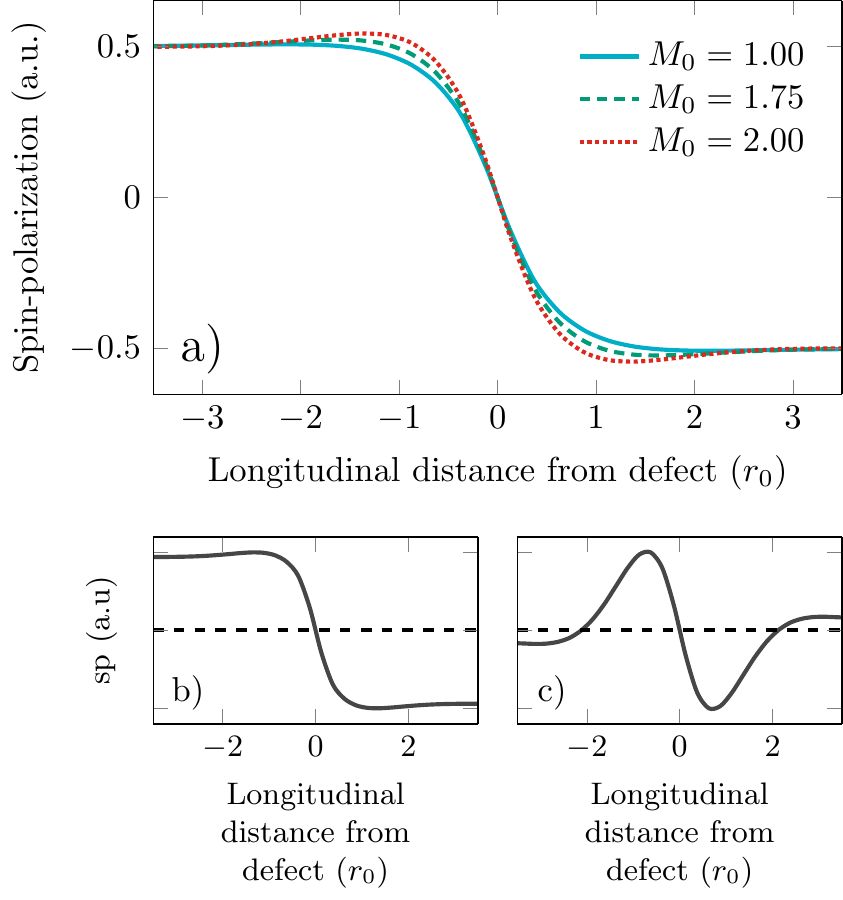}
	\caption{Contributions to the spin-density redistribution from an un-polarized magnetic defect for $M_0 = 1.50, 1.75, 2.00 E_0$ in blue solid, green dashed, and red dotted respectively. The primary effect is to redistribute equal proportions of spin in either direction, however a secondary ringing effect is observed based on the value of $M_0$. At $M_0<A_0^2/B_0$, the secondary effect is negative, causing a smoother redistribution. However at $M_0>A_0^2/B_0$ it becomes additive, causing a sharp ringing feature to appear, 3 examples of which are illustrated here. The extrema of primary and secondary features are at approximately $\pm 1.5r_0$ and $\pm 0.75r_0$ respectively. This ringing effect is examined in Fig.~\ref{fig:dIseparation}.}
	\label{fig:probespin}
\end{figure}

All the results so far can be understood in terms of the standard Boltzmann transport picture \cite{Bozkurt2018,Lunde2012}. However, an interesting effect seen in Figs.~\ref{fig:varyM}, \ref{fig:reinjected}, and \ref{fig:probespin} is that although the impurity is modelled as point-like, the redistributed spin-current density exhibits finite extent before reaching its steady-state behaviour, typically on the length scale of 1--3 $r_0$. To examine these effects we computed the spin-resolved current density as a function of distance from the defect \emph{in the direction of transport}, a subset of which are shown in Fig.~\ref{fig:probespin}a. Using principle component analysis on 9 samples spaced across $M_0=0.5 E_0$ to $M_0=2.5 E_0$ we observe two primary contributions, one corresponding to the underlying current redistribution independent of $M_0$ (Fig \ref{fig:probespin}b) and another undulating factor which depends of the value of $M_0$ (Fig.~\ref{fig:probespin}c). We note that all biases studied are well within the bulk gap and as such are unaffected by any band modifications about the bandgap edges; the effect depends solely on the states inside the gap.

\begin{figure}\centering
	\includegraphics[]{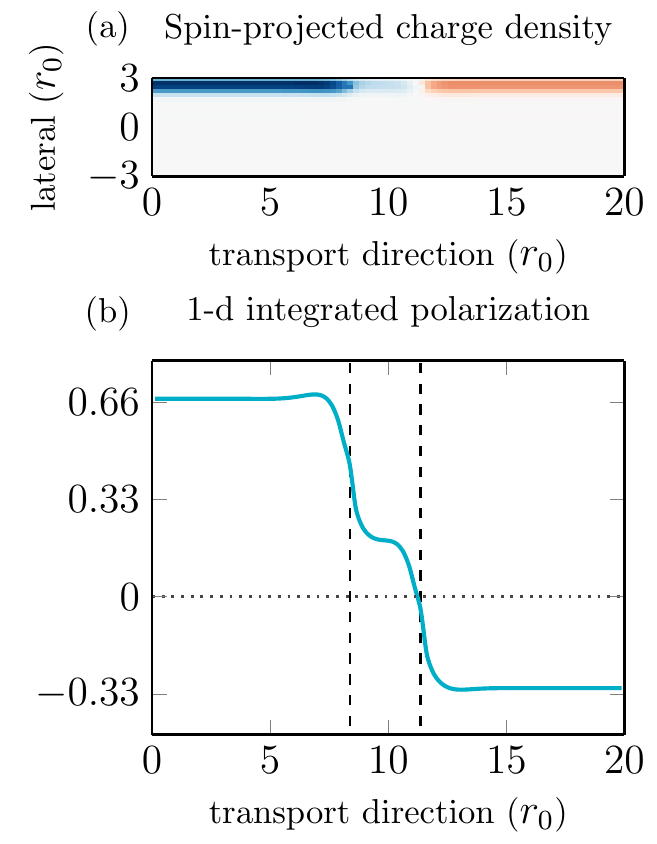} 
	\caption{ (a) Spin-projected density of states for combined impurities both with $\langle m \rangle = 0.5$. (b) One-dimensionally integrated spin-projected DOS, normalized by the difference between the maximum and minimum spin contributions. Dotted line cuts through zero (\textit{i.e.} no change of the HES due to impurity scattering) and dashed lines cut through the defect locations. }
	\label{fig:twoprobes}
\end{figure}

Now we use the same formalism to consider two defects positioned within a HES to examine how the finite extent of the HES response modifies the device response of high defect densities. We show the purely additive effect of two defects placed 4$r_0$ apart in Fig.~\ref{fig:twoprobes}. 

At higher densities we find that the effects are not purely additive. In order to determine the contributions of two-defect effects we note that different polarizations, $M_0$ values and spatial positions all cause differing conductance changes, and as such we introduce the following ratio describing the relative difference between two isolated and non-isolated defects:
\begin{equation}\label{ratioG}
\text{Conductance decrease ratio} = \frac{d\sigma_\text{1+2} - d\sigma_\text{1} - d\sigma_\text{2} }{d\sigma_\text{1} + d\sigma_\text{2}},
\end{equation}
where $d\sigma_{1+2}$ is the total conductance decrease from two impurities, and $d\sigma_{1}, d\sigma_{2}$ are the individual conductance decreases from each impurity when the other is not present. If there are no higher order effects, then this contribution would be exactly zero, and if the contribution doubled due to higher order effects, this value would be exactly unity.

\begin{figure}\centering
	\includegraphics[]{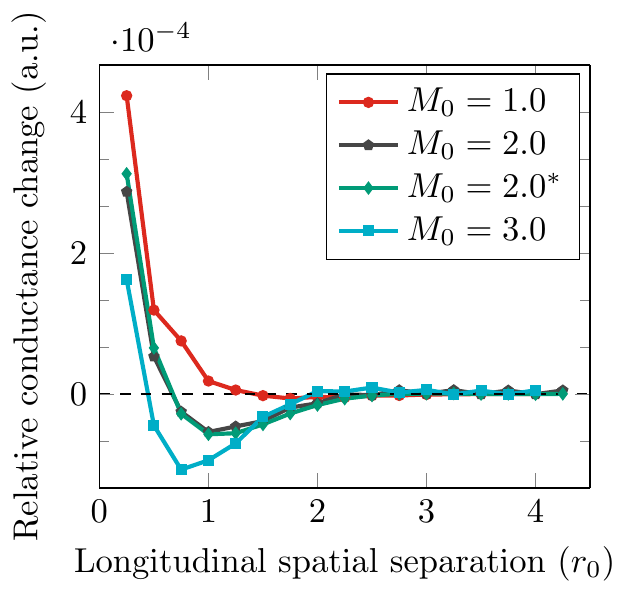}
	\caption{Relative change in conductance of two unpolarized ($\langle m \rangle = 0$) spin impurities, for (red circles) $M_0=1.0E_0$, (grey pentagons) $M_0=2.0E_0$ and (blue squares) $M_0=3.0E_0$, also the effect of two partially polarized impurities ($\langle m \rangle = 0.5$) is shown in (green diamonds) for $M_0=2.0E_0$ denoted by asterisk. Positive indicates that the device is less conductive than two isolated defects, whereas negative indicates it is more conductive. Here we can see that transitioning across differing values of $M_0$, the device transitions from being less to more conductive. The results also appear to be somewhat insensitive to the polarization of the defects. We qualitatively observe the sinc-like behaviour in the relative conductance change but at a much smaller scale.} 
	\label{fig:dIseparation}
\end{figure}

These results are presented in Fig.~\ref{fig:dIseparation} where defects are linearly separated in units of $0.25 r_0$ (the lattice spacing). We see that the influence on conduction possesses both negative and positive contributions, that is, more and less conductive regions and that its minimum depends on the choice of $M_0$, but not the polarization, $\langle m \rangle$, as demonstrated by the overlapping curves with fixed $M_0=2$, but $\langle m \rangle = 0.0, 0.5$. It appears that the $M_0$-dependent effects are maximized for impurity separations close to the edge state length scale $r_0$ and that the relative conductance change is negligible for separations $\gtrsim4 r_0$, implying a maximum density of scatterers beyond which the effects due to multiple scatterers are not simply additive.

The non-additive effect of multiple defects on conductance can be understood in terms of spin-flip scattering discussed in Section \ref{spinflip}. For a single defect $N=1$, the conductance is obtained $\delta \sigma_1/G_0 = k_\text{f}\delta T$ following Eqs. (\ref{sfG}) and (\ref{eq:deltaT}), where ${k_\text{f}}={E_\text{f}/(\hbar v_0)}$. 
Similarly, for double defects $N=2$, we have:
\begin{align}
\delta \sigma_2/G_0 = 2k_\text{f}\delta T+ 2k_\text{f} \sinc(2 k_\text{f} l)\delta T,
\end{align}
where $\sinc(x) = \sin(x)/x$. In this case, the conductance decrease oscillates with $k_\text{f}$ and decays with the defect separation $l$. At large $l$, it reduces to $\delta \sigma_2=2k_\text{f} \delta T=2\delta \sigma_1$, implying an additive effect of the decoupled defects. Now the ratio in Eq. (\ref{ratioG}) is explicitly derived as
\begin{equation}
\frac{\delta \sigma_2 - 2 \delta \sigma_1}{2 \delta \sigma_1 } = \sinc(2 k_\text{f} l) 
\end{equation}
in this limit, if we plot this ratio we observe sinc-like behaviour, which is consistent with our numerical results depicted in Fig.~\ref{fig:dIseparation}. The characteristic length scale of the sinc oscillation is given by $k_\text{f}$ 
\begin{equation}
k_\text{f} \leadsto \frac{E_0}{\hbar v_0} = \frac{|A_0|^2/2|B_0| }{\sqrt{B_0^2-D_0^2} |A_0|/|B_0| } \stackrel{D_0\rightarrow 0}{=} \frac{|A_0|}{2 |B_0|} \equiv \frac{1}{r_0}.
\end{equation}
which is consistent with $1/r_0$ observed in the numerical results. Although the function form matches well, the magnitude of the effect is not comparable between the effective and NEGF models. However this is not surprising as this derivation is for a single electron and one-dimensional scattering, and the results in Fig.~\ref{fig:dIseparation} are for an ensemble of electrons in 2D. 

For a general number of defects $N$, we can obtain an approximation of the conductance decrease as
\begin{align}
\delta \sigma_N/G_0\approx& N k_\text{f} \delta T + 2 (N-1) k_\text{f} \sinc(2 k_\text{f} l) \delta T, \label{eq:Ndefects}
\end{align}
where we discarded the higher order oscillation terms. The total conductance decrease scales with the number of defects as expected.

\section{Conclusion}
Using the NEGF formalism we have developed a model for understanding the role of magnetic impurities in the BHZ model. We have shown that a fully microscopic model agrees qualitatively with the semi-classical Boltzmann treatment for single impurities. However a secondary effect is observed between two impurities depending on the material parameter $M_0$. This effect modifies the response at high impurity densities, implying a maximum density of magnetic impurities beyond which a more sophisticated treatment is required. 

Using a real-space Green's function approach allows us to consider spatially dependent effects and realistic device geometries. These methods may be applied to studies in different parameter regimes and for different materials, as our results are quite general. The approach can also be applied to model the effects of non-point-like defects such as charge puddles, as well as spin-charge correlated disorder~\cite{Keser:PRL2019}, by adapting both the in-scattering function and spatial distribution of the local potential within the device. Studying the influence of spin-dependent scattering on topological edge states due to magnetic impurities is important to understand the limitations of spin-momentum locked states for use in energy storage and spintronics.

\section{Acknowledgements}
We acknowledge useful discussions with S. Wilkinson, D. Culcer, M. Fuhrer and B. Muralidharan. This work was supported in part by the Australian Research Council under the Centre of Excellence funding scheme CE170100039. Computational resources were provided by the NCI National Facility systems at the Australian National University through the National Computational Merit Allocation Scheme supported by the Australian Government. This research was supported by a Lockheed Martin Corporation Research Grant.

\bibliography{bhzbib}

\appendix

\section{Numerical Hilbert Transform} \label{App:Hilbert}
Computation of the Hilbert transform is typically considered a numerically expensive process, here we show that it may be reduced to a simple matrix multiplication, following the work of \cite{Shishkin2006}. However, unlike \cite{Shishkin2006} which was designed for a different purpose, we do not require our functions to be symmetric in energy and instead provide a more general solution. We begin by noting that the real and imaginary components of causal functions must be Hilbert transforms of each other:
\begin{align}
\text{Re}\left[\Sigma(E)\right] =& -\frac{1}{\pi} \int_{-\infty}^{\infty}  \frac{\text{Im}\left[\Sigma(E^\prime)\right]}{E-E^\prime}dE^\prime \label{eq:hilbtrans} \\
=& \text{Im}\left[\Sigma(E) \right] \otimes \left( \frac{ -1}{\pi E} \right).
\end{align}
Equation (\ref{eq:hilbtrans}) may be transformed to a discrete energy grid by approximating the kernel with energy-dependent weights multiplied by some basis function $\Phi(E)$:
\begin{equation}
\Sigma(E) \approx \sum_{j} \Sigma(E_j) \Phi_j(E)
\end{equation}
here $E$ is a continuous variable, whereas $E_j$ are discrete grid points. Following this, one may compute the Hilbert transform of the basis functions evaluated at the target energy point $E_k$:
\begin{equation}
\phi_{kj} = \phi(E_k,E_j) = \int_{-\infty}^{\infty} dE^\prime \Phi_j(E^\prime) \left( \frac{1}{E_k - E^\prime - i 0^+} \right),
\end{equation}
where $0^+$ is a small infinitesimal to provide stability to the integration. In our work $\Phi_i(E^\prime)$ are a collection of triangular shaped functions centred at $E_i$, which serve analogously to the delta function on a discrete grid (or, for another interpretation are linearly interpolating functions):
\begin{equation}
\Phi_j(E^\prime) = \left\{\begin{array}{lr}
\dfrac{E^\prime - E_{j-1}}{E_j - E_{j-1}}, & \text{for } E_{j-1} \leq E^\prime \leq E_j \\
\dfrac{E_{j+1} - E^\prime}{E_{j+1} - E_{j}}, & \text{for } E_{j} \leq E^\prime \leq E_{j+1} \\
0, & \text{elsewhere}
\end{array}\right.  
\end{equation}
due to the definition of our triangle functions, these points do not need to be placed on an equidistant grid, allowing us prioritize sampling depending on our requirements. These integrals can be formed analytically (and thus the broadening term may be omitted) to give:
\begin{multline}
\phi_{kj} = \frac{E_k-E_{j-1}}{E_{j}-E_{j-1}}\log \left|\dfrac{E_{j-1}-E_k}{E_{j}-E_k}\right| \\  +\dfrac{E_k-E_{j+1} }{E_{j}-E_{j+1}}\log \left|\dfrac{E_k-E_{j}}{E_k-E_{j+1}}\right| 
\end{multline}
where $\log|x|$ is the natural logarithm of the absolute value of $x$. With these weights, Hilbert transforms such as this one can now be performed by simple summation:
\begin{equation}
\text{Re}[G_{lm}(E_k)] = -\frac{1}{\pi} \sum_{j} \phi_{kj} \text{Im}[ G_{lm}(E_j)]
\end{equation}
as we are on a discrete grid, energies lower and higher than certain points will not be calculated, as such it is important to make sure that the functions being evaluated have sufficiently converged to zero. Evaluation of the Hilbert transform matrix need only be computed once, and all subsequent transforms are found solely by matrix multiplication.

\begin{figure}\centering
	\includegraphics[]{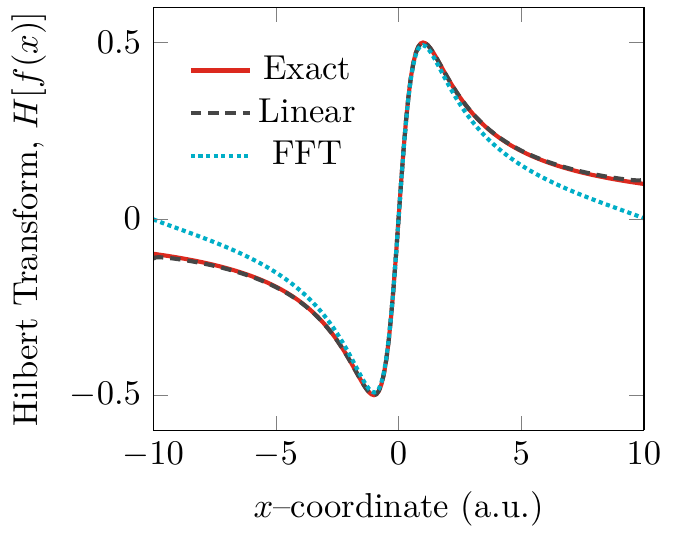} 
	\caption{Comparison of numerical Hilbert transforms of $f(x)=$  $1/(x^2+1)$. (red, solid) Exact analytic transform, $ H[f(x)]=$ $x/(x^2+1)$ (black, dashed) our linear basis function method, and (blue, dotted) FFT method. Linear in this sense refers to the interpolation of the measured data and not of the transform. Note because of the periodic assumption, the FFT method goes to zero at the edges. A slight deviation at the edges of our method may be observed due to truncating the original function before it reached zero. This may be corrected by expanding the window.} 
	\label{fig:Hilb}
\end{figure}

Other authors \cite{Lemus2020} have chosen to compute the Hilbert transform using the Fast Fourier Transform (FFT) approach. The FFT method relies on the definition of the convolution operator being a product in Fourier space. One first computes the FFT of the signal and of $-1/E$ and then inverse transforms them back. This method though computationally inexpensive has significant drawbacks. Firstly, because it uses FFT, the function is inherently assumed to be periodic. This forces the positive and negative solutions to be clamped to zero requiring needlessly large window sizes to achieve accurate results. Secondly, \textit{prima facie} it does not support non-uniform grids, though efficient non-uniform alternatives to the FFT have been proposed \cite{Potter2017} based upon band-limited interpolation. A comparison of FFT and our method is shown in Fig.~\ref{fig:Hilb}.

\section{Additional numerical methods}

\subsection{Numerical integration}
To compute numerical integration of observables we use a shape-preserving piecewise cubic Hermite interpolating polynomial using MATLAB's \verb|PCHIP| function, then analytically integrate it using \verb|fnint|. We found that this method has the benefits of Simpson's integration formulae but maintains physicality (non-negative density of states) with the added flexibility of \mbox{non-uniform} grids.

\subsection{Self-consistency of impurity self-energies}
To aid in the convergence of self-energies we use Eyert's modified broyden method \cite{Eyert1996}. For a linear mixing of the input and output of some function $f$, (in this case a linearized version of our self-energy $f=\text{vec}(\Sigma^r)$), this iteration may be written as:
\begin{equation}
f_\text{in}^{i+1} = (1-\alpha) f_\text{in}^{i} +  \alpha f_\text{out}^{i}
\end{equation}
where $f_\text{in}^{i}$and $f_\text{out}^{i}$ were the input and output states for the $i^\text{th}$ iteration, $\alpha$ is the linear mixing parameter and $f_\text{in}^{i+1}$ is the new input guess. Defining the difference in output states as the ``function'' \mbox{$F^{i} = f_\text{out}-f_\text{in}$}, and $X^{i} = f_\text{in}$, and the Jacobian inverse $J^{-1} = -\alpha I$, we arrive at a quasi-Newton iteration formula
\begin{equation}
X^{i+1} = X^{i} - J^{-1} F^{i}
\end{equation}
and therefore may use the techniques of Eyert's modified Broyden scheme discussed below. Unlike conventional schemes which perform a rank-one update from the previous iteration, Eyert's method performs a minimal \mbox{rank-$n$} update in the least-squares sense for several previous steps simultaneously generating a much better approximation to the underlying Jacobian. Given a starting guess for the \textit{inverse} Jacobian \mbox{$J^{-1}\approx G_0$}, the new updated inverse Jacobian $G$, is given by
\begin{equation}
G = G_0 + (dx - G_0 dF) W
\end{equation}
$dx$ is a matrix whose columns are spanned by the difference in input guesses $dx_i = X^{i} - X^{i-1}$, and $dF$ similarly is a matrix spanned by the difference in outputs $dF_i = F^{i} - F^{i-1}$, and $W$ is a projection matrix given by:
\begin{equation}
W = \begin{cases}
\text{pinv}(dF), & \text{``Bad''}, \\
(Z^\dagger  dF )\backslash Z^\dagger, & \text{``Good''}.
\end{cases}
\end{equation}
where $Z = G_0 dx$. In this work we chose $\alpha = 0.75$, and 6 Broyden steps to be included (though typically we achieve convergence after 1--2 steps. In Eyert's work, his projection matrix was based on the ``Bad'' Broyden rank-one update, for our work we similarly define what we call the ``Good'' projection matrix that we based on the ``Good'' Broyden rank-one update by inspection.

\subsection{Active update of $G^n$}
The self-consistent cycle requires both the iteration of $G^r$ and $G^n$ together. $G^n$ and its Hilbert transform enter into the self-consistent solution of $G^r$, but as the Hilbert transform is non-local property, it cannot be updated until all energy points have been computed, the common practice is then to keep $G^n$ completely fixed until all $G^r$ are computed. Instead, we fix only the Hilbert transform of $G^n$, and actively update the anti-Hermitian part (typically by linear mixing with the old solution), the anti-Hermitian part corresponds to the density of occupied states and allows for a more active feedback. Though this may slightly increase the computation time, the \textit{additional} computation of $G^n$ is cheap compared to $G^r$, as it requires only matrix products ($N^2$) rather than inverses ($N^3$), and all the relevant set-up costs are in the initial computation of $G^r$; we find that this method leads to fewer outer ($G^n$) iterations overall. However, as noted in the following subsection, for a large collection of points, the conditional computation of $G^n$ reduces this additional computational cost to zero.

\subsection{Conditional computation of $G^n$} \label{app:condGn}
In our work we apply a conditional computation of $G^n$ for our self-consistent cycle, consider the Fermi-Dirac distribution function:
\begin{equation}
f(E) = \left( \exp\left( \frac{E-\mu}{kT} \right) + 1 \right)^{-1},
\end{equation}
we consider that for energies lower than the lowest chemical potential, all state filling (including those from impurity self-energies) may be well described by Fermi-Dirac distribution functions; at energies much lower than the lowest chemical potential, the Fermi-Dirac distribution function can be safely approximated by:
\begin{equation}
f(E) \approx \begin{cases}
1 - \exp\left( \frac{E-\mu}{kT} \right), & \text{for } E \ll \mu, \\
\exp\left( \frac{\mu - E}{kT} \right), & \text{for } E \gg \mu.
\end{cases}
\end{equation}
Then, for energies much lower than the lowest chemical potential, the in-scattering matrix:
\begin{equation}
\Sigma^{\text{in}} = \sum_{i}^{N} f_i \Gamma_i
\end{equation}
may safely be approximated by $\Gamma$:
\begin{equation}
\Gamma= \sum_{i}^{N} \Gamma_i.
\end{equation}
as $f_i \approx 1$. Upon this substitution, one notes that 
\begin{equation}
G^n = G^r \Sigma^\text{in} G^a \approx G^r \Gamma G^a = A
\end{equation}
and computation of $G^n$ is unnecessary as it is equal to the anti-hermitian part of $G^r$. If the approximation is made at $E = \mu - pkT$, then the largest error is $e^{-p}$. If $p \geq 36.84$ then the answer will be indistinguishable from floating point imprecision. However, we have found that perhaps due to the use of products to compute $G^n$, when we examined the differences between $A$ and $G^n$ numerically, we found that at $p = 18.42$ the results became indistinguishable. Note, this does not remove the necessity of the Hilbert transform in these regions as it is a non-local property.

\section{Spin-flip induced magnetization}\label{sfM}
For a single spin defect and conduction electron pair, the eigenstate is given by:
\begin{equation}
\ket{\psi_c}=  c_{\uparrow \downarrow}\ket{\! \!\uparrow \downarrow}+ c_{\downarrow \uparrow} \ket{ \! \! \downarrow \uparrow}.
\end{equation}
Here, $c_{\uparrow \downarrow} $ and $c_{\uparrow \downarrow}$ are:
\begin{gather}
c_{\uparrow \downarrow}=\left( c_1\cos\frac{\theta}{2} e^{iqx}+c_2\sin\frac{\theta}{2} e^{-iqx} \right), \\
c_{\downarrow \uparrow}= \left( c_1\sin\frac{\theta}{2} e^{iqx}+c_2\cos\frac{\theta}{2} e^{-iqx} \right),
\end{gather}
where the coefficients $c_{1,2}$ are 
\begin{gather}
c_1=\frac{\sec\frac{\theta}{2}}{1-e^{2iqw}\tan^2{\frac{\theta}{2}}},\\
c_2=\frac{e^{2iqw}\sin\frac{\theta}{2}}{-\cos^2{\frac{\theta}{2}}+e^{2iqw}\sin^2{\frac{\theta}{2}}}.
\end{gather}
The full density matrix of electron-defect system is $\rho_\text{tot}=|\psi_c\rangle\langle\psi_c|$, from which the reduced density matrix of the defect spin is obtained as $\rho=\text{Tr}_e(\rho_\text{tot})$, where the trace is over electron spin space. Explicitly, 
\begin{eqnarray}
\rho=\left(\begin{array}{cc}
F_u&0\\
0&F_d\end{array}\right),
\end{eqnarray}
with $F_u=\left|c_{\downarrow \uparrow}\right|^2, F_d=\left|c_{\uparrow \downarrow}\right|^2$.
The diagonal elements of the defect density matrix are given by
\begin{gather}
F_u = \frac{\sin ^2(\theta) \sin ^2(q (w-x))}{\cos ^2(\theta) \cos ^2(q w)+\sin ^2(q w)},\\
F_d = \frac{3 - 2 \sin ^2(\theta) \cos (2 q (w-x))+\cos (2 \theta) }{3 - 2 \sin ^2(\theta) \cos (2 q w)+\cos (2 \theta)}.
\end{gather}

\end{document}